\documentclass[11pt]{article}
\pdfoutput=1
\usepackage{geometry}
\usepackage{cite,subfigure}
\usepackage{amsmath}
\usepackage{amssymb}
\usepackage{amsfonts}
\usepackage{graphicx}
\usepackage{caption}
\usepackage{float}
\usepackage[usenames,dvipsnames]{xcolor}
\usepackage{tikz}
\usepackage{pstricks}
\usepackage{authblk}
\usepackage{verbatim}
\usepackage{setspace}
\usepackage[pagebackref=true]{hyperref}
\hypersetup{
	colorlinks=true,
	linkcolor=blue,
	filecolor=Violet,     
	urlcolor=RubineRed,
	citecolor=red
}
\begin{document}
\date{}
\author{Rajibul Shaikh \thanks{\href{mailto:rshaikh@iitk.ac.in}{rshaikh@iitk.ac.in}}}
\author{Pritam Banerjee \thanks{\href{mailto:bpritam@iitk.ac.in}{bpritam@iitk.ac.in}}}
\author{Suvankar Paul \thanks{\href{mailto:svnke@iitk.ac.in}{svnkr@iitk.ac.in}}}
\author{Tapobrata Sarkar \thanks{\href{mailto:tapo@iitk.ac.in}{tapo@iitk.ac.in}}}
\affil{Department of Physics, \\ Indian Institute of Technology, \\ Kanpur 208016, India}
\title{Strong gravitational lensing by wormholes}
\maketitle

\begin{abstract}
We study strong gravitational lensing by a class of static wormhole geometries. Analytical approaches to the same
are developed, and the results differ substantially from strong lensing by black holes, first reported by Bozza. We consider two
distinct situations, one in which the observer and the source are on the same side of the wormhole throat, and
the other in which they are on opposite sides. Distinctive features in our study arise from the fact that photon
and antiphoton spheres might be present on both sides of the wormhole throat, and that the throat might itself act as a photon sphere.
We show that strong gravitational lensing thus opens up a rich variety of possibilities of relativistic image formation,
some of which are novel, and are qualitatively distinct from black hole lensing. These can serve as 
clear indicators of exotic wormhole geometries. 

\end{abstract}

\maketitle

\section{Introduction}

Gravitational lensing is known to be one of the most important tools to test the predictions of general
relativity (GR). Weak lensing from celestial objects has been extensively studied over the last century, while 
lensing in the strong gravity limit is a relatively recent area of research that has received much attention 
of late \cite{SL1,SL2,SL3,SL4,Tsu1,WGL1,WGL2,WGL3}. In the light of the first results coming from the Event Horizon Telescope \cite{EHT1,EHT2,EHT3}, it is of great interest
to further these lines of research. 

Important in this context are horizonless structures, which are fast gaining popularity as potentially important
objects to distinguish from the ubiquitous black hole backgrounds 
\cite{UCO1,UCO2,UCO3,UCO4,UCO5,UCO_MNRAS,UCO_aps}. One of the prime reasons for this is that these can 
mimic black hole lensing effects \cite{UCO2,UCO_MNRAS}. One important class of horizonless objects are wormhole 
backgrounds\footnote{We consider only symmetric wormholes in this paper}, where two
distinct universes or two distant regions of the same universe are connected by a throat region \cite{morris1,visser_book}. 
Such geometries were introduced following Einstein and Rosen's
proposal of the Einstein-Rosen bridge \cite{ER_bridge}. Later, 
the term {\em wormhole} was first introduced by Misner and Wheeler \cite{misner1}. 
Although wormhole models in GR require the presence of exotic matter, it is now known that, in modified gravity theories, these can exist 
with normal matter as well \cite{WMG1,WMG2,WMG3,WMG4,WMG5,WMG6,WMG7,
WMG8,WMG9,WMG10,WMG11,WMG12,WMG13,WMG14,WMG15,WMG16,WMG17,WMG18,WMG19,
WMG20,WMG21,WMG22,WMG23,WMG24,WMG25,WMG26,WMG27,WMG28,WMG29,WMG30}. 

Various aspects of gravitational lensing by wormholes in both the weak and the strong deflection limit have been 
studied in the literature and have been compared to that by black holes \cite{WL0,WL1,WL2,WL3,WL4,
WL5,WL6,WL7,WL8,WL9,WL10,WL11,WL12,Tsu2,WL14,WL15,WL16,WL17,WL18,WL19,
WL20,WL21,WS1,WS2,WS3,WS4,WS5,WS6}. Important 
in the context of strong lensing are photon and antiphoton spheres, which are characterized by radii at which photons can have
unstable and stable circular orbits, respectively. Strong lensing traditionally refers to photons that are trapped at an
unstable photon surface and undergo multiple rotations until a small perturbation can make them escape to
infinity. Antiphoton spheres also play a crucial role when they exist, as has been recently shown in \cite{UCO_aps}. 
Strong field lensing in a general spherically symmetric and static spacetime was first studied analytically 
by Bozza in \cite{SL3} and later by Tsukamoto in \cite{Tsu1}. Besides being used to study strong lensing by black holes, 
the analytic methods developed in \cite{SL3,Tsu1} have also been used in most of the studies on strong gravitational 
lensing by wormholes available in the literature. 

The broad purpose of this paper is two-fold. First, 
we point out that the formulas obtained in \cite{SL3,Tsu1} to study strong lensing by wormholes may fail in some 
cases when the wormhole throat acts as an effective photon sphere, and we develop alternate analytic formulas to 
address strong lensing in such cases. Moreover, almost all the earlier studies on the strong gravitational lensing by 
wormholes have been in the scenario where the light source and the observer are on the same side of the wormhole throat. 
Since the throat of a wormhole connects two distant asymptotic regions, there can be another lensing scenario where 
the light source and the observer are on the opposite sides. Gravitational lensing in a particular example, namely 
the Ellis-Bronnikov wormhole in this latter scenario has been addressed in \cite{WLOS1,WLOS2}. 
In this paper, we develop the analytic formalism for strong gravitational lensing by a general spherically symmetric 
and static wormhole geometry with the source and the observer on opposite sides of the throat.

The broad outcome of this paper is that the presence of multiple photon and antiphoton spheres results in a rich 
and novel structure of relativistic images formed due to the strong gravitational lensing by wormholes than the ones 
that have been reported previously in the literature. In particular, as discussed above, the following two distinct
scenarios might arise in the physics of lensing from generic static, spherically symmetric wormholes:
\begin{itemize}
\item
{\bf a.} The observer and the source are on the same side of the wormhole throat. Three sub-cases might
arise here: \\
{\bf a1.} Strong lensing occurs due to the presence of a photon sphere outside the throat. \\
{\bf a2.} The throat itself acts as a photon sphere in strong lensing. \\
{\bf a3.} Strong lensing occurs due to both a photon sphere and an antiphoton sphere. 
\item
{\bf b.} The observer and the source are on opposite sides of the throat, i.e, the observer sees light coming
from another universe. Two distinct cases  can occur here, which are formally similar to the cases {\bf a1} and 
{\bf a2}, i.e, \\
{\bf b1.} A photon sphere outside the throat is involved in strong lensing. \\
{\bf b2.} The throat itself acts as a photon sphere in strong lensing. 
\end{itemize}
These are studied analytically in this paper, for generic wormhole geometries. 

Indeed, the discussion above opens up a variety of new possibilities in strong lensing by wormholes. 
In the next section, we recapitulate the essential features of symmetric wormhole geometries,
and elaborate on the strong lensing possibilities discussed above. In section \ref{sec:sameside}, we study 
strong lensing of light in case {\bf a}, i.e, when the observer and the source are on the same side of the throat. 
Then, in section \ref{sec:otherside}, we consider situation {\bf b}, when the observer and the source are on opposite
sides of the wormhole throat. Section \ref{sec:examples} contains three specific examples that
exemplify the computation of the previous sections. In section \ref{sec:observables}, we list the phenomenological
observables in gravitational lensing and provide numerical results. Finally, we conclude this paper in section
\ref{sec:conclusions} with some discussions on our results. Appendix \ref{sec:Appendix} contains the
details of the case {\bf b2} which is not included in the main text, for ease of reading.


\section{Wormholes and the deflection angle of light}
\label{sec:deflection}

We consider a general spherically symmetric, static wormhole of the Morris-Thorne class, whose 
spacetime metric can be written in spherical polar coordinates as \cite{morris1}
\begin{equation}
ds^2=-e^{2\Phi(r)} dt^2+\frac{dr^2}{1-\frac{\mathcal{B}(r)}{r}}+r^2(d\theta ^2+\sin ^2\theta d\phi ^2)~.
\label{eq:metric_r}
\end{equation}
Here $\Phi(r)$ and $\mathcal{B}(r)$ are called the redshift function and the wormhole shape function, respectively. 
The wormhole throat specifies the connection between two different regions, and is given by  
$\left(1-\frac{\mathcal{B}(r)}{r}\right)\Big\vert_{r_{th}}=0$, i.e., by $\mathcal{B}(r_{th})=r_{th}$, with $r_{th}$ 
being the radius of the throat. $\mathcal{B}(r)$ satisfies the flare-out condition $\mathcal{B}'(r_{th})<1$ also \cite{morris1}. 
Note that $\Phi(r)$ must be finite everywhere (from the throat to spatial infinity).

If we define a proper radial coordinate $l(r)$ as 
\begin{equation}
l(r)=\pm\int_{r_{th}}^r\frac{dr}{\sqrt{1-\frac{\mathcal{B}}{r}}},
\label{eq:proper_radial}
\end{equation}
in terms of which the throat is at $l(r_{th})=0$, and the two signs (plus and minus) correspond to the two different regions 
connected through the throat, then we can write the line element of Eq. (\ref{eq:metric_r}) as 
\begin{equation}
ds^2=-e^{2\Phi(l)} dt^2+dl^2+r^2(l)(d\theta ^2+\sin ^2\theta d\phi ^2).
\label{eq:metric_l}
\end{equation}
In general, however, it might be difficult to invert the relation in Eq. (\ref{eq:proper_radial}) to obtain $r(l)$ in which case 
it will not be possible to write the wormhole metric explicitly in terms of the proper radial coordinate $l$.\footnote{Note that, in the proper radial coordinates, 
the wormhole throat is now given by (using Eq. (\ref{eq:proper_radial}))
$
r'(l)\Big\vert_{l=0}=\left(1-\frac{\mathcal{B}(r)}{r}\right)\Big\vert_{r=r_{th}}=0,
$
where the prime denotes a differentiation with respect to its argument. }

For convenience and ease of notation, we work with the generic static metric given by
\begin{equation}
ds^{2}=-A(r)dt^{2}+B(r)dr^{2}+C(r)(d\theta^{2}+\sin^{2}\theta d\phi^{2})~,
\label{genmetric}
\end{equation}
and specific wormhole examples will be worked out later. 
Also, we assume that the wormhole is symmetric with respect to its throat and is asymptotically flat. 
Therefore, the metric functions satisfy the asymptotically flat conditions
\begin{eqnarray}
\lim_{r\rightarrow \infty} A(r) = 1~,~\lim_{r\rightarrow \infty} B(r) = 1~,~\lim_{r\rightarrow \infty} C(r) = r^{2}.
\end{eqnarray}
Because of the spherical symmetry, we can choose $\theta=\pi/2$. With this choice, the 
Lagrangian corresponding to the motion of photons in the background geometry of the wormhole represented by 
Eq. (\ref{genmetric}) is
\begin{equation}
2\mathcal{L}=-A(r)\dot{t}^2+B(r)\dot{r}^2+ C(r) \dot{\phi}^2,
\end{equation}
where an overdot represents a derivative with respect to the affine
parameter. Since there are two Killing vectors $\partial_t$ and $\partial_{\phi}$, there are two constants of motion, namely
\begin{equation}
p_t=\frac{\partial\mathcal{L}}{\partial\dot{t}}=-A(r)\dot{t}=-E~,~
p_\phi=\frac{\partial\mathcal{L}}{\partial\dot{\phi}}=C(r)\dot{\phi}=L,
\end{equation}
where $E$ and $L$ are, respectively, the energy and angular momentum
of the photon. From the normalization condition
$g_{\mu\nu}{u}^\mu{u}^\nu=0$, one obtains
\begin{equation}
AB\dot{r}^2+V_{eff}=E^2, \hspace{0.3cm}
V_{eff}={L^2}\frac{A(r)}{C(r)},
\label{Veffec}
\end{equation}
where $V_{eff}$ is the effective potential. The impact parameter of a light ray (which remains constant throughout the trajectory 
of a photon) is defined as $b=L/E$. Depending on the effective potential, a 
photon coming from a source at infinity may turn at some radius $r_0$ 
and then escapes to a faraway observer. Such a turning point is indicated by $\dot{r}=0$, i.e., by $V_{eff}(r_0)=E^2$. 
This implies that 
\begin{equation}
b^2=\frac{C(r_0)}{A(r_0)}.
\label{eq:impact_parameter}
\end{equation}
For a photon which comes from a distant source, takes a turn at $r_0$ and escapes to a faraway observer, 
the deflection angle $\alpha(r_{0})$ is given by the well known formula \cite{Weinberg,Virbhadra:1998dy,Claudel:2000yi,Virbhadra:2002ju}
\begin{equation}
\alpha(r_{0})=I(r_{0})-\pi,
\label{eq:deflection1}
\end{equation}
where we have defined
\begin{equation}
I(r_{0})= 2\int^{\infty}_{r_{0}}\frac{dr}{\sqrt{\frac{R(r)C(r)}{B(r)}}}~,~
R(r)= \left(\frac{A_{0}C}{AC_{0}}-1 \right).
\label{eq:I1}
\end{equation}
Strong gravitational lensing occurs when $r_0$ is close to the location of the photon sphere, i.e., a radius at which 
light can bend in angles excess of $2\pi$. The photon sphere comprises unstable photon orbits. In general, there might be stable 
photon orbits as well, which constitute an antiphoton sphere.
Circular photon orbits satisfy $V_{eff}=E^2$ and $dV_{eff}/dr=0$, resulting in Eq. (\ref{eq:impact_parameter}) and
\begin{equation}
\frac{C'(r)}{C(r)}-\frac{A'(r)}{A(r)}=0~,
\label{eq:psph}
\end{equation}
respectively \cite{Weinberg,Virbhadra:1998dy,Claudel:2000yi,Virbhadra:2002ju}. In addition, at the location of a photon and an antiphoton sphere, we must have, 
respectively, $d^2V_{eff}/dr^2<0$ (maximum of the potential) and $d^2V_{eff}/dr^2>0$ (minimum of the potential). 
In this paper, the position of the photon sphere is denoted by $r = r_m$, and the corresponding critical impact 
parameter as $b = b_m=\sqrt{C(r_m)/A(r_m)}$. Equation (\ref{eq:psph}) is satisfied at $r=r_m$.

As we have shown in \cite{WL_novel}, in addition to the above-mentioned photon and antiphoton spheres located 
outside the throat, for a wormhole which is symmetric with respect to its throat, the throat itself can act as a position of either a 
maximum (effective photon sphere) or a minimum (effective antiphoton sphere) of the effective potential. 
However, this may not be true for an asymmetric wormhole \cite{bronnikov_asymmetric} in general. In the following, 
when the throat acts as an effective photon sphere, we denote the corresponding critical impact parameter by 
$b_{th}=\sqrt{C(r_{th})/A(r_{th})}$.

We now systematically address below all the different cases of strong lensing mentioned in the Introduction.

\section{Strong lensing of light when the observer and the source are on the same side of the throat}
\label{sec:sameside}

We begin by considering the situations where the observer and the source are on the same side of the throat. 
As mentioned in the Introduction, there are three distinct subcases here. 

\subsection{Case {\bf a1} : Strong bending of light due to a photon sphere outside the throat}
\label{subsec:sameside_a}

\begin{figure}[ht]
\centering
\subfigure[]{\includegraphics[scale=0.55]{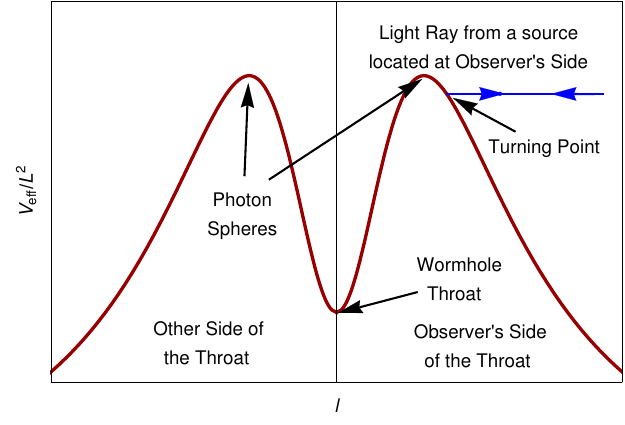}}
\subfigure[]{\includegraphics[scale=0.55]{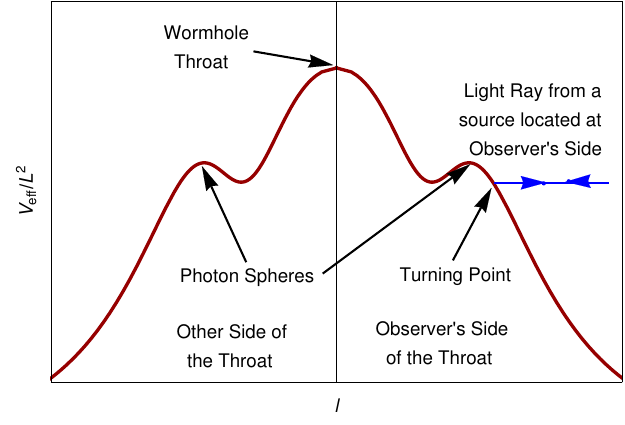}}
\subfigure[]{\includegraphics[scale=0.55]{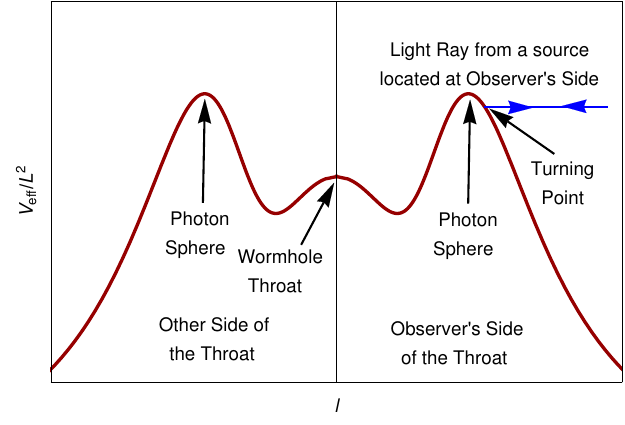}}
\caption{Case {\bf a1 :} Schematic diagrams showing strong lensing of light due to a photon sphere outside the throat in a wormhole spacetime. The maxima of the effective potential (in units of angular momentum squared) outside the throat represent the positions of photon spheres.}
\label{fig1}
\end{figure}

Let us now consider strong lensing of light due to a photon sphere located outside the throat, in the scenario when 
the observer and the light source are on the same side of the throat. This is case {\bf a1} illustrated in Fig. \ref{fig1}. 
A light ray with an impact parameter $b$ greater than the critical value $b_m$ always takes a turn outside the photon 
sphere and escapes to the faraway observer. The strong deflection in this case occurs in the limit 
$r_{0}\to r_{m}$ or $b\to b_m$ ($b\geq b_m$). This strong deflection of light due to the photon sphere located 
outside the wormhole throat is similar to the one for a black hole studied in detail in \cite{SL3,Tsu1}. In this 
case, the deflection angle in the strong deflection limit comes out to be \cite{Tsu1}
\begin{equation}
\alpha(b)=-\bar{a}\log \left( \frac{b}{b_m}-1 \right) +\bar{b} +\mathcal{O}\left[\left(b-b_m\right)\log\left(b-b_m\right)\right],
\label{eq:strong_alpha_1}
\end{equation}
where $\bar{a}$ and $\bar{b}$ are given by
\begin{equation}
\bar{a}=\sqrt{\frac{2B_{m}A_{m}}{C^{''}_{m}A_{m}-C_{m}A^{''}_{m}}}~,~
\bar{b}=\bar{a}\log \left[r^{2}_{m}\left(\frac{C_{m}^{''}}{C_{m}}-\frac{A_{m}^{''}}{A_{m}}\right)\right] +I_{R}(r_{m})-\pi,
\label{eq:strong_bbar_1}
\end{equation}
respectively, with the subscript `$m$' implying the corresponding quantities evaluated at $r=r_m$. Note that, since the strong lensing in this case occurs when the impact parameter $b$ approaches the critical value $b_m$ from $b>b_m$ side, the resulting relativistic images are formed at impact parameters greater than the critical value $b_m$, i.e., they are formed just outside the photon sphere.


\subsection{Case {\bf a2} : Strong bending of light due to a wormhole throat}
\label{subsec:sameside_b}

\begin{figure}[ht]
\centering
\subfigure[]{\includegraphics[scale=0.80]{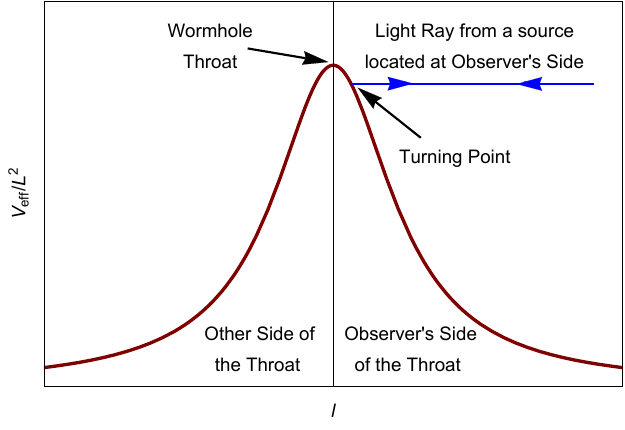}}
\subfigure[]{\includegraphics[scale=0.80]{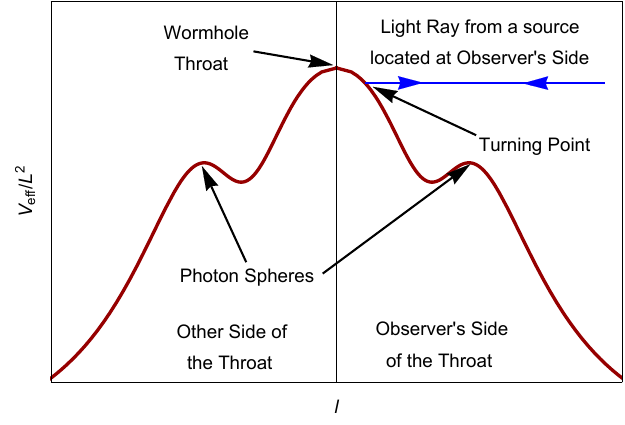}}
\caption{Case {\bf a2 :} Schematic diagrams showing strong lensing of light due to a wormhole throat which acts as the maximum of the effective potential (in units of angular momentum squared).}
\label{fig3}
\end{figure}

We now come to situation {\bf a2}, where 
the wormhole throat itself acts as an effective photon sphere, i.e., as a position of the maximum of the effective potential for photons. 
As a result, light can experience strong lensing due to the throat. To realize such strong lensing due to the throat,
it is not necessary to have another maximum of the potential outside the throat to mark the position of a photon sphere. 
If there is one outside the throat, then the height of the maximum at the throat must be greater than that at the outer photon 
sphere to realize the strong lensing due to the throat. Figure \ref{fig3} schematically illustrates this situation. 

In this case, a photon with impact parameter $b>b_{th}$ always has a turning point outside the throat, and the 
strong deflection limit occurs when the turning point approaches the throat, i.e., $ r_0\to r_{th} $ or the impact 
parameter approaches the critical value $b_{th}$, i.e., $ b\to b_{th} $ from $ b>b_{th} $ side. 
Photons with impact parameters $b<b_{th}$ get captured by the throat and escape to the other side. 
Here we would like to mention that, to obtain the strong deflection angle in this case, we can still use Eqs. (\ref{eq:strong_alpha_1})-(\ref{eq:strong_bbar_1}) with $r_m=r_{th}$ and $b_m=b_{th}$ only when the wormhole 
spacetime can be written in the proper radial coordinates so that $B_m=B(r_m)=B(r_{th})$ does not diverge. 
See Sec. III(C) of \cite{Tsu1} for such an example. However, in many cases, it is difficult to express the wormhole 
metric explicitly in the proper radial coordinates, and we have to deal with the one expressed in spherical polar 
coordinates given in (\ref{eq:metric_r}). If this is the case, then we cannot use 
Eqs. (\ref{eq:strong_alpha_1})-(\ref{eq:strong_bbar_1}) to obtain the strong deflection angle due to the throat as 
the metric function $B(r)=1/\left(1-\mathcal{B}(r)/r\right)$ diverges at the throat. Therefore, in this case, we need to 
develop a formula of light bending in the strong deflection limit. 

To this end, we introduce a variable $z$ defined as 
\begin{equation}
z= 1-\frac{r_0}{r}.
\end{equation}
Putting this in $I(r_{0})$ in Eq.(\ref{eq:I1}), we obtain
\begin{equation}
I(r_{0})=\int^{1}_{0}f(z,r_0)dz,
\end{equation}
where
\begin{equation}
f(z,r_0)= \frac{2r_0}{\sqrt{G(z,r_0)}}~,~
G(z,r_0)= R\frac{C}{B}(1-z)^{4}.
\label{defG0}
\end{equation}
Since $ B(r_{0})\to \infty $ in the strong deflection limit $ r_0\to r_{th} $, we define a new variable as $ \bar{B}(r)= 1/B(r)=\left(1-\mathcal{B}(r)/r\right) $ so that $ \bar{B}(r_{th})=0 $.
Therefore, we have
\begin{equation}
G(z,r_0)= RC\bar{B}(1-z)^{4}.
\label{defG}
\end{equation}
Now, we need to expand $G(z,r_0)$ near $r=r_{0}$ or $z=0$ to extract its divergent part. Here, it should be noted that the expansion of a function $F(r)$ in powers of $z$ around $ z=0 $ can be written as
\begin{equation}
F=F_0+F_0^{'}r_{0}z+\left(\frac{1}{2}F^{''}_{0}r_0^{2}+F_0^{'}r_0 \right)z^{2}+\mathcal{O}(z^{3})\nonumber .
\end{equation}
In the above expression as well as in the rest of this discussion, the subscript `$ 0 $' indicates that the quantities are evaluated at $r=r_{0}$ or $z=0$.
Therefore, the expansion of $R(r)$ in the powers of $z$ can be written as
\begin{equation}
R(r)= r_0 \left(\frac{C_0^{'}}{C_0}-\frac{A_0^{'}}{A_{0}}\right)z 
+\left[\frac{r_0^2}{2} \left(\frac{C_{0}^{''}}{C_{0}}-\frac{A_{0}^{''}}{A_{0}}\right)+r_0\left(1-\frac{A_0^{'}r_0}{A_{0}}\right)\left(\frac{C_0^{'}}{C_0}-\frac{A_0^{'}}{A_{0}}\right)\right]z^2 +\mathcal{O}(z^3)
\end{equation}
Similarly, expanding the functions $\bar{B}$ and $C$ in powers of $z$ in Eq. (\ref{defG}), we get the full expansion of $G(z,r_0)$ as
\begin{equation}
G(z,r_0)=\delta z+\eta z^2+\mathcal{O}(z^3),
\end{equation}
where we have defined
\begin{equation}
\delta=r_0 C_{0} \bar{B}_{0} \left(\frac{C_0^{'}}{C_0}-\frac{A_0^{'}}{A_{0}}\right)
\label{deltaint}
\end{equation}
\begin{eqnarray}
\eta = r_0\left(\frac{C_0^{'}}{C_0}-\frac{A_0^{'}}{A_{0}}\right)\left[r_0\left(C_0\bar{B}_0^{'}+C_0^{'}\bar{B}_0\right)-C_0\bar{B}_0\left(3+\frac{A_0^{'}r_0}{A_0}\right)\right]  +
\frac{r_0^2}{2}C_{0}\bar{B}_{0}\left(\frac{C_{0}^{''}}{C_{0}}-\frac{A_{0}^{''}}{A_{0}}\right)~.
\label{etaint}
\end{eqnarray}

Once again, note that the function $ \bar{B}(r) $ vanishes at $ r=r_{th} $, i.e., $ \bar{B}(r_{th})=0 $. Therefore, in the limit $r_0 \to r_{th}$, we obtain
\begin{equation}
\delta_{th}=\delta\vert_{r_0=r_{th}}=0, \quad \eta_{th}=\eta\vert_{r_0=r_{th}}=r_{th}^2 C_{th}\bar{B}_{th}^{'}\left(\frac{C_{th}^{'}}{C_{th}}-\frac{A_{th}^{'}}{A_{th}}\right),
\end{equation}
where the subscript `$th$' indicates that the quantities are evaluated at $r=r_{th}$. Hence, we get
\begin{equation}
G_{th}(z)=\eta_{th} z^{2}+\mathcal{O}(z^{3}).
\end{equation}
This shows that the leading order of the divergence of $f(z, r_0)$ is $z^{-1}$ and that the integral $I(r_{0})$ diverges logarithmically in the strong deflection limit $r_0\rightarrow r_{th}$, as was the case for black holes in \cite{SL3,Tsu1}.

To extract out the logarithmic divergence part, the integral $I(r_0)$ is split into two parts -- a divergent part 
$I_D(r_0)$ and a regular part $I_R(r_0)$, such that $I(r_{0})=I_{D}(r_{0})+I_{R}(r_{0})$. 
The divergent part $I_D(r_0)$ is defined as
\begin{equation}
I_D(r_0)= \int^{1}_{0}f_D(z,r_0)dz~, \quad
f_D(z,r_0)
=\frac{2r_0}{\sqrt{\delta z+\eta z^2}}.
\end{equation}
The regular part $I_R(r_0)$, on the other hand, is defined as
\begin{equation}
I_{R}(r_{0})=\int^{1}_{0} f_R(z,r_0)dz~, \quad
f_R(z,r_0)=f(z,r_0)-f_D(z,r_0).
\end{equation}
Integrating $I_D(r_0)$, we get 
\begin{equation}
I_D(r_0)=\frac{4r_0}{\sqrt{\eta}}\log \frac{\sqrt{\eta}+\sqrt{\delta+\eta}}{\sqrt{\delta}}.
\end{equation}
After doing some algebra,  we obtain in the limit $r_0\to r_{th}$, 
\begin{equation}
I_D(r_0)=-\frac{2r_{th}}{\sqrt{\eta_{th}}}\log(r_{0}-r_{th}) + \frac{2r_{th}}{\sqrt{\eta_{th}}}\log(4r_{th}) + \mathcal{O}\left[(r_0-r_{th})\log(r_{0}-r_{th})\right],
\label{IDr0}
\end{equation}
where the following expansion has been used:
\begin{equation}
\bar{B}_0 = \bar{B}_{th}^{'}(r_0-r_{th})+\mathcal{O}(r_0-r_{th})^2.
\end{equation}
Moreover, we can also write
\begin{eqnarray}
b^2=\frac{C_0}{A_0}=\left[\frac{C_{th}+C_{th}^{'}(r_0-r_{th})+\mathcal{O}(r_0-r_{th})^2}{A_{th}+A_{th}^{'}(r_0-r_{th})+\mathcal{O}(r_0-r_{th})^2}\right]
= b_{th}^2\left[1+\left(\frac{C_{th}^{'}}{C_{th}}-\frac{A_{th}^{'}}{A_{th}}\right)(r_0-r_{th})\right]+\mathcal{O}(r_0-r_{th})^2
\label{rat}
\end{eqnarray}
Note that $b\to b_{th}$ when $r_0\to r_{th}$. Therefore, using the last expression, the divergent part $I_{D}(b)$ in terms of the impact parameter $b$ in the strong deflection limit $b\to b_{th}$ takes the form
\begin{eqnarray}
I_D(b)
=-\frac{2r_{th}}{\sqrt{\eta_{th}}} \log \left( \frac{b^2}{b_{th}^2}-1 \right) +\frac{2r_{th}}{\sqrt{\eta_{th}}} \log \left[4r_{th}\left(\frac{C_{th}^{'}}{C_{th}}-\frac{A_{th}^{'}}{A_{th}}\right)\right] +\mathcal{O}[(b^2-b_{th}^2)\log (b^2-b_{th}^2)].
\end{eqnarray}

Similarly, if we expand the regular part $I_R(r_0)$ too in powers of ($r_0-r_{th}$) in the strong deflection limit, keeping the leading order term only, we obtain
\begin{equation}
I_R(r_0)= \int^{1}_{0} f_R(z,r_{th})dz+\mathcal{O}((r_0-r_{th})\log(r_0-r_{th}))
\end{equation}
which can be expressed in terms of the impact parameter as
\begin{equation}
I_R(b)= \int^{1}_{0} f_R(z,b_{th})dz+\mathcal{O}((b^2-b_{th}^2)\log (b^2-b_{th}^2)).
\end{equation}
Therefore, we finally obtain the bending angle of light in the strong deflection limit $r_0\to r_{th}$ or $b\to b_{th}$ due to the wormhole throat as
\begin{equation}
\alpha(b)=-\bar{a}\log \left( \frac{b^2}{b_{th}^2}-1 \right) +\bar{b} +\mathcal{O}((b^2-b_{th}^2)\log(b^2-b_{th}^2)),
\label{eq:strong_alpha_3}
\end{equation}
where
\begin{equation}
\bar{a}=2\sqrt{\frac{A_{th}}{\bar{B}_{th}^{'}(C^{'}_{th} A_{th}-C_{th} A^{'}_{th})}}, \quad \bar{b}=\bar{a}\log \left[4r_{th}\left(\frac{C_{th}^{'}}{C_{th}}-\frac{A_{th}^{'}}{A_{th}}\right)\right] +I_R(r_{th})-\pi.
\label{eq:strong_bbar_3}
\end{equation}
The expressions in Eqs. (\ref{eq:strong_alpha_3})-(\ref{eq:strong_bbar_3}) are quite different from the corresponding expressions in the previous case due to a photon sphere located outside the throat. Note that the above formula is valid when the wormhole metric is written in spherical polar coordinate so that $ \bar{B}(r)= 1/B(r)=\left(1-\mathcal{B}(r)/r\right) $ vanishes at the throat. In order that the dependence of the above analytic formula on $b$ looks similar to the ones due to a photon sphere outside the throat (Eqs. (\ref{eq:strong_alpha_1})-(\ref{eq:strong_bbar_1})), we use the approximation $\left( \frac{b^2}{b_{th}^2}-1 \right)\simeq 2 \left( \frac{b}{b_{th}}-1 \right)$ in the strong deflection limit $b\to b_{th}$ and obtain
\begin{equation}
\alpha(b)=-\bar{a}\log \left( \frac{b}{b_{th}}-1 \right) +\bar{b} +\mathcal{O}((b-b_{th})\log(b-b_{th})),
\label{eq:strong_alpha_3a}
\end{equation}
where
\begin{equation}
\bar{a}=2\sqrt{\frac{A_{th}}{\bar{B}_{th}^{'}(C^{'}_{th} A_{th}-C_{th} A^{'}_{th})}}, \quad \bar{b}=\bar{a}\log \left[2r_{th}\left(\frac{C_{th}^{'}}{C_{th}}-\frac{A_{th}^{'}}{A_{th}}\right)\right] +I_R(r_{th})-\pi.
\label{eq:strong_bbar_3a}
\end{equation}
Note the change in $\bar{b}$ compared to the one in eq. (\ref{eq:strong_bbar_3}). Note also that, since the strong lensing in this case occurs when the impact parameter $b$ approaches the critical value $b_{th}$ from $b>b_{th}$ side, the resulting relativistic images are formed at impact parameters greater than the critical value $b_{th}$, i.e., they are formed just outside the throat.

\subsection{Case {\bf a3} : Strong bending of light experiencing an antiphoton sphere}
\label{subsec:sameside_c}

\begin{figure}[ht]
\centering
\includegraphics[scale=0.80]{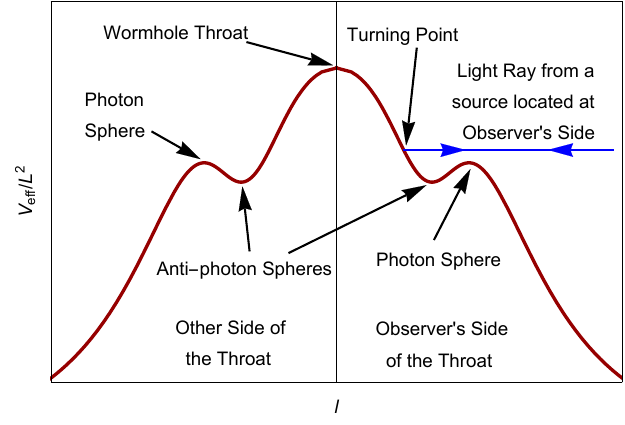}
\caption{Case {\bf a3 :} Schematic diagram showing strong lensing of light due to an antiphoton sphere in a wormhole spacetime. The throat acts as a maximum of the effective potential (in units of angular momentum squared) in addition to another maximum (photon sphere) outside of it. The minimum of the effective potential in between this two maxima represents the position of the antiphoton sphere.}
\label{fig2}
\end{figure}

We now consider the case when the light ray encounters an antiphoton sphere. If the effective potential exhibits a maximum at the throat $ r=r_{th} $ (effective photon sphere) as well as at $ r=r_m $ (photon sphere) with $ r_m>r_{th} $, then there will be a minimum (antiphoton sphere) at $ r=r_{aps} $ (say) in between the two maxima. This minimum of the effective potential acts as the location of stable circular orbits of photons. If the height of the maximum of the effective potential at the throat is greater than that at the outer photon sphere, a photon having an impact parameter greater than $b_{th}$ but less than $b_m$ enters both the photon and the antiphoton spheres, takes a turn at a radius inside the antiphoton sphere, and comes out of the photon sphere and escapes to a
faraway observer. See Fig. \ref{fig2} for illustration. In this case, the strong deflection occurs when the impact parameter approaches the critical value $b_m$ from $b < b_m$ side. This case is similar to the strong lensing due to the presence of an antiphoton sphere around an ultracompact object discussed in \cite{UCO_aps}. The strong deflection angle of light in the limit $b\to b_m$ ($b\leq b_m$) in this case is given by \cite{UCO_aps}

\begin{equation}
\alpha(b)=-\bar{a}\log \left( \frac{b_m^2}{b^2}-1 \right) +\bar{b} +\mathcal{O}\left[\left(b_m^2-b^2\right)\log\left(b_m^2-b^2\right)\right],
\label{eq:strong_alpha_2}
\end{equation}
where
\begin{equation}
\bar{a}=2\sqrt{\frac{2B_m A_m}{C^{''}_m A_m-C_m A^{''}_m}}, \quad \bar{b}=\bar{a}\log \left[2r_m^2\left(\frac{C_m^{''}}{C_m}-\frac{A_m^{''}}{A_m}\right)\left(\frac{r_m}{r_c}-1\right)\right] +I_R(r_c)-\pi,
\label{eq:strong_bbar_2}
\end{equation}
and $r_c$ is the radius at which the effective potential has the same height as that at the photon sphere $r=r_m$. See \cite{UCO_aps} for more details. Note that, since the strong lensing in this case occurs when the impact parameter $b$ approaches the critical value $b_m$ from $b<b_m$ side, the resulting relativistic images are formed at impact parameters less than the critical value $b_m$, i.e., they are formed just inside the photon sphere.


\section{Strong lensing of light coming from the other side of a wormhole throat}
\label{sec:otherside}

We now study the scenario when the light source and the observer are located on the opposite sides of the wormhole throat. So, the light rays have to cross the throat as well as the photon and anti-photon spheres located in between to reach the faraway observer. Therefore, in this scenario, we must always have $V_{eff}(r)<E^2$ along the photon geodesics, and hence, there is no turning point $r_0$. The requirement $V_{eff}(r)<E^2$ can be achieved by suitable choices of $E$ and $L$ or, equivalently, by a suitable choice of the impact parameter $b$ ($=L/E$). Note that, since the photon directly passes through the throat without having any turning point, unlike in the scenario discussed in the previous section, here the impact parameter of light cannot be expressed as a function of the turning point. As a result, the bending angle of light ($ \alpha $) can also not be expressed in terms of $r_0$. Instead, all the quantities are expressed in terms of the impact parameter $ b $. Therefore, the fact that there is no turning point is going to be crucial while studying gravitational lensing. The deflection angle $\alpha(b)$ of the light in this case can be written as
\begin{equation}
\alpha(b)=I(b)-\pi,
\end{equation}
where
\begin{equation}
I(b)= 2\int^{\infty}_{r_{th}}\frac{dr}{\sqrt{\frac{R(r)C(r)}{B(r)}}}~, \quad
R(r)=\left( \frac{C}{b^2 A}-1\right).
\label{eq:alpha_b}
\end{equation}
Like the scenario discussed in the previous section, here also we assume that both the observer and the source are far away 
from the throat. We also assume that the observer is at $\phi=0$ on one side of the throat, and the source is at around 
$\phi=\pi$ on the other side. Below, we discuss two different cases separately that can arise here.

\subsection{Case {\bf b1} : Strong bending of light due to a photon sphere outside the throat}
\label{subsec:otherside_a}

\begin{figure}[ht]
\centering
\subfigure[]{\includegraphics[scale=0.80]{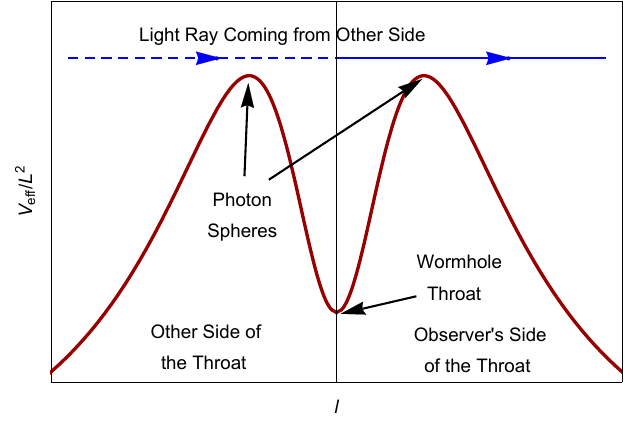}}
\subfigure[]{\includegraphics[scale=0.80]{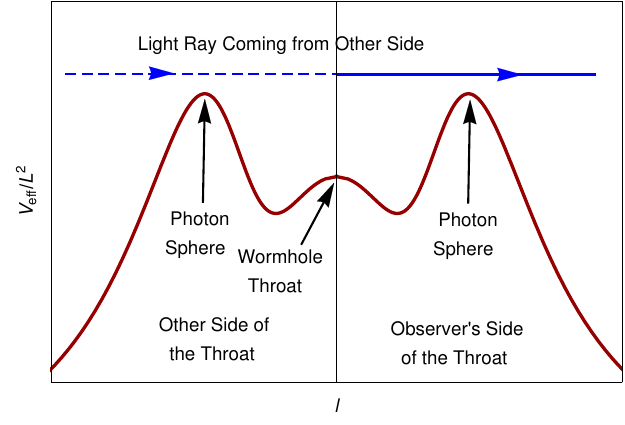}}
\caption{Case {\bf b1 :} Schematic diagrams showing strong lensing of light due to a photon sphere outside the throat in a wormhole spacetime. The maxima of the effective potential (in units of angular momentum squared) outside the throat represent the positions of photon spheres. In this case, light comes from the other side, passes through the throat and then reaches to the observer.}
\label{fig4}
\end{figure}

Here, we shall discuss strong bending of light due to the presence of a photon sphere outside the wormhole throat, when the light from the source located on the other side crosses the throat and reaches the observer (see Fig. \ref{fig4}). Note that there may be multiple photon spheres on the two sides of the throat. Let the photon sphere corresponding to the highest maximum of the effective potential be 
located at $ r=r_m $ and have the critical impact parameter $b_m$. Photons from the light source located on the other side and with impact parameter $b>b_m$ always get turned away to the same side and do not reach the observer. On the other hand, photons with impact parameter $b< b_m$ cross the photon sphere as well as the throat and reach the faraway observer. The strong deflection limit in this case occurs when the impact parameter $b$ approaches the critical value $b_m$, i.e., $ b\to b_{m} $ from $ b<b_{m} $ side.

To obtain the strong deflection angle in this case, we define
\begin{equation}
z= 1-\frac{r_{m}}{r}~.
\end{equation}
Using this in (\ref{eq:alpha_b}), we obtain
\begin{equation}
I(b)=\int^{1}_{1-\frac{r_{m}}{r_{th}}}f(z,b,r_m)dz,
\end{equation}
where
\begin{equation}
f(z,b,r_m)= \frac{2r_{m}}{\sqrt{G(z,b,r_m)}}~, \quad
G(z,b,r_m)= R\frac{C}{B}(1-z)^{4}.
\end{equation}
Following the steps discussed in the previous section, $R(r)$ in this case can be expanded in powers of $z$ as
\begin{equation}
R(r)=\left(\frac{C_m}{A_m b^2}-1\right) 
+\frac{r_{m}^2}{2} \frac{C_m}{b^2 A_m} \left(\frac{C_{m}^{''}}{C_{m}}-\frac{A_{m}^{''}}{A_{m}}\right)z^2 +\mathcal{O}(z^3).
\end{equation}
Using similar expansions of $ C $ and $ B $, we obtain the expansion of $G(z,b,r_m)$ in powers of $z$ as 
\begin{equation}
G(z,b,r_m)=\gamma+\delta z+\eta z^2+\mathcal{O}(z^3),
\end{equation}
where
\begin{equation}
\gamma=\frac{C_m}{B_m}\left(\frac{C_m}{A_m b^2}-1\right),
\end{equation}
\begin{equation}
\delta=\frac{C_m}{B_m}\left(\frac{C_m}{A_m b^2}-1\right)\left[-4+r_m\left(\frac{C_{m}^{'}}{C_{m}}-\frac{B_{m}^{'}}{B_{m}} \right)\right],
\end{equation}

\begin{eqnarray}
\eta &=&\frac{C_m}{B_m}\left(\frac{C_m}{A_m b^2}-1\right)\left[6-r_m\left(3+\frac{B_m^{'}r_m}{B_m}\right)\left(\frac{C_{m}^{'}}{C_{m}}-\frac{B_{m}^{'}}{B_{m}} \right)\right. \nonumber \\
& & \left.  +\frac{r_m^2}{2}\left(\frac{C_{m}^{''}}{C_{m}}-\frac{B_{m}^{''}}{B_{m}}\right)\right]+\frac{r_m^2}{2}\frac{C_m}{B_m}\frac{C_m}{A_m b^2}\left(\frac{C_{m}^{''}}{C_{m}}-\frac{A_{m}^{''}}{A_{m}}\right).
\end{eqnarray}
Note that, since $\frac{C(r_m)}{A(r_m)}=b_m^2$, we have $\left(\frac{C_m}{A_m b^2}-1\right)\to 0$ in the limit $b \to b_m$. Therefore, in this limit, we get
\begin{equation}
\gamma_m=\gamma\vert_{b=b_m}=0=\delta_m=\delta\vert_{b=b_m}, \quad \eta_m=\eta\vert_{b=b_m}=\frac{r_m^2}{2}\frac{C_m}{B_m}\left(\frac{C_{m}^{''}}{C_{m}}-\frac{A_{m}^{''}}{A_{m}}\right).
\end{equation}
Hence, we obtain
\begin{equation}
G_{m}(z)=\eta_m z^{2}+\mathcal{O}(z^{3}).
\end{equation}
Just like the cases discussed before, the leading order of the divergence of $f(z, b,r_m)$ in this case too goes as $z^{-1}$ and so the integral $I(b)$ diverges logarithmically in the strong deflection limit $b\to b_m$.

To extract out this divergent part in the strong deflection limit, we need to split up the integral $I(b)$ into a divergent part $I_D(b)$ and a regular part $I_R(b)$, i.e., $I(b)=I_{D}(b)+I_{R}(b)$. Now, the divergent part $I_D(b)$ is defined as
\begin{equation}
I_D(b)= \int^{1}_{1-\frac{r_m}{r_{th}}}f_D(z,b,r_m)dz, \quad f_D(z,b,r_m)
=\frac{2r_m}{\sqrt{\gamma+\delta z+\eta z^2}}.
\label{eq:Id(b)_int}
\end{equation}
Whereas, the regular part $I_R(b)$ is defined as
\begin{equation}
I_{R}(b)=\int^{1}_{1-\frac{r_m}{r_{th}}} f_R(z,b,r_m)dz, \quad f_R(z,b,r_m)=f(z,b,r_m)-f_D(z,b,r_m).
\end{equation}
Performing the integration in Eq. (\ref{eq:Id(b)_int}), we obtain 
\begin{equation}
I_D(b)=\frac{2r_m}{\sqrt{\eta}}\log \frac{\delta+2\eta+2\sqrt{\eta}\sqrt{\gamma+\delta+\eta}}{\delta+2\eta\left(1-\frac{r_m}{r_{th}}\right)+2\sqrt{\eta}\sqrt{\gamma+\delta\left(1-\frac{r_m}{r_{th}}\right)+\eta\left(1-\frac{r_m}{r_{th}}\right)^2}}.
\end{equation}
In the limit $b\to b_m$, we get
\begin{equation}
I_D(b)=\frac{2r_m}{\sqrt{\eta_m}}\log\left[\frac{4\eta_m\left(\frac{r_m}{r_{th}}-1\right)}{\frac{C_m}{B_m}\left(\frac{b_m^2}{b^2}-1\right)}\right]+\mathcal{O}\left[\left(\frac{b_m^2}{b^2}-1\right)\log\left(\frac{b_m^2}{b^2}-1\right)\right].
\end{equation}
Again, the regular part $I_R(b)$, after expanding in powers of $b_m^2-b^2$ and keeping the leading order term, can be written in integral form as,
\begin{equation}
I_R(b)= \int^{1}_{1-\frac{r_m}{r_{th}}} f_R(z,b_m,r_m)dz+\mathcal{O}((b_m^2-b^2)\log(b_m^2-b^2))
\end{equation}
Finally, the bending angle of light in the strong deflection limit $b\to b_m$ ($ b\leq b_m $) can be written as
\begin{equation}
\alpha(b)=-\bar{a}\log \left( \frac{b_m^2}{b^2}-1 \right) +\bar{b} +\mathcal{O}((b_m^2-b^2)\log(b_m^2-b^2)),
\label{eq:strong_alpha_4}
\end{equation}
where
\begin{equation}
\bar{a}=2\sqrt{\frac{2B_m A_m}{C^{''}_m A_m-C_m A^{''}_m}}, \quad \bar{b}=\bar{a}\log \left[2r_m^2\left(\frac{C_m^{''}}{C_m}-\frac{A_m^{''}}{A_m}\right)\left(\frac{r_m}{r_{th}}-1\right)\right] +I_R(b_m)-\pi.
\label{eq:strong_bbar_4}
\end{equation}
Note that the above expressions are similar with those in Eqs. (\ref{eq:strong_alpha_2}-\ref{eq:strong_bbar_2}), except that $r_c$ is now replaced by $r_{th}$. Note also that, since the strong lensing in this case occurs when the impact parameter $b$ approaches the critical value $b_m$ from $b<b_m$ side, the resulting relativistic images are formed at impact parameters less than the critical value $b_m$, i.e., they are formed just inside the photon sphere.


\subsection{Case {\bf b2} : Strong bending of light due to a wormhole throat}
\label{subsec:otherside_b}

\begin{figure}[ht]
\centering
\subfigure[]{\includegraphics[scale=0.80]{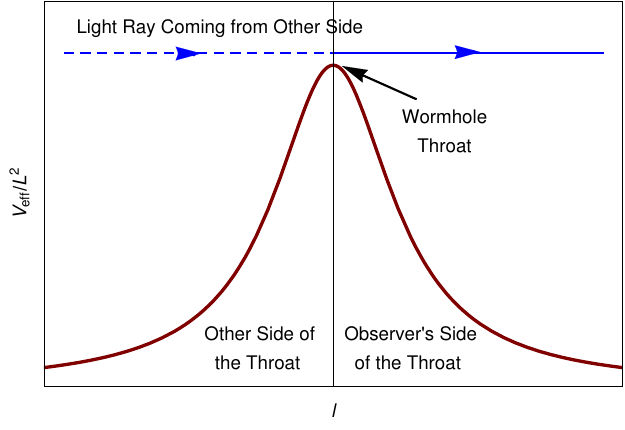}}
\subfigure[]{\includegraphics[scale=0.80]{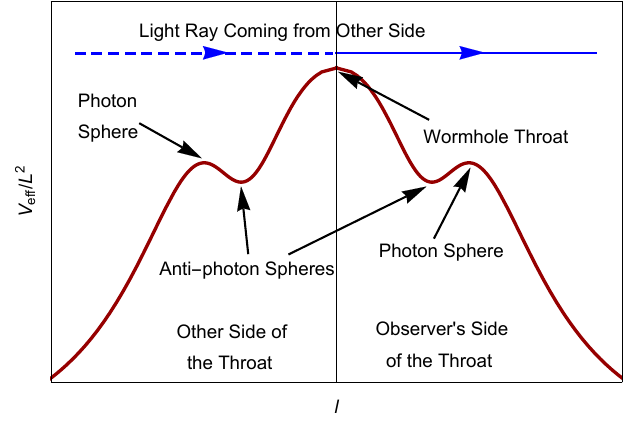}}
\caption{Case {\bf b2 :} Schematic diagrams showing strong lensing of light due to a wormhole throat which acts as the maximum of the effective potential (in units of angular momentum squared). In this case also, light comes from the other side, passes through the throat and then reaches to the observe.}
\label{fig5}
\end{figure}

We now discuss strong lensing of light when the wormhole throat acts as an effective photon sphere (see Fig. \ref{fig5}). 
In this case, photons which are from the source on the other side and have impact parameter $b>b_{th}$ always have 
turning point outside the throat on the same side. These photons do not reach the observer and have marginal 
turning point $r_{th}$ when $b=b_{th}$. Photons with $b<b_{th}$ do not have any turning point, cross the throat and reach the observer. 
The strong deflection limit in this case occurs in the limit $ b\to b_{th} $ from $ b<b_{th} $ side. For the same reason 
(same as the one discussed in subsection \ref{subsec:sameside_b}) that $B_m=B(r_m)=B(r_{th})$ diverges at the throat, 
we cannot use Eqs. (\ref{eq:strong_alpha_4})-(\ref{eq:strong_bbar_4}) to obtain the strong deflection angle in this case when 
the spacetime metric is written in spherical polar coordinate. Therefore, we need to rederive the expression for the strong 
deflection angle in this case.

The methodology is similar to the previous cases, and for ease of reading, the details of the calculation in this case
are relegated to Appendix \ref{sec:Appendix}. We only quote the final result given in Eq. (\ref{eq:strong_alpha_5})
for the deflection angle in the strong deflection limit $b\to b_{th}$ ($b\leq b_{th}$) as
\begin{equation}
\alpha(b)=-\bar{a}\log \left( \frac{b_{th}^2}{b^2}-1 \right) +\bar{b} + \mathcal{O}\left[\left(b_{th}^2-b^2\right)\log \left(b_{th}^2-b^2\right)\right],
\label{eq:strong_alpha_5_app}
\end{equation}
where
\begin{equation}
\bar{a}=2\sqrt{\frac{A_{th}}{\bar{B}^{'}_{th}\left(C^{'}_{th}A_{th}-A^{'}_{th}C_{th}\right)}}, \quad \bar{b}=\bar{a}\log \left[4r_{th}\left(\frac{C^{'}_{th}}{C_{th}}-\frac{A^{'}_{th}}{A_{th}}\right)\right] +I_R(b_{th})-\pi.
\label{eq:strong_bbar_5}
\end{equation}
Note that, since the strong lensing in this case occurs when the impact parameter $b$ approaches the 
critical value $b_{th}$ from $b<b_{th}$ side, the resulting relativistic images are formed at impact parameters less 
than the critical value $b_{th}$, i.e., they are formed just inside the throat. It is also interesting to note that the expressions 
in Eqs. (\ref{eq:strong_alpha_5_app})-(\ref{eq:strong_bbar_5}) exactly match with the corresponding expressions in 
Eqs. (\ref{eq:strong_alpha_3})-(\ref{eq:strong_bbar_3}), except that the dependence of the logarithmic term on $b$ 
and $b_{th}$ is somewhat different in the two cases. Both of these expressions correspond to gravitational lensing 
in the strong deflection limit due to the wormhole throat when the throat acts as the maximum of the effective potential. 

As a result of this, as we shall see in Sec. \ref{sec:observables}, the relativistic images which are formed just outside 
the throat ($ b>b_{th} $) and those formed just inside the throat ($ b<b_{th} $) will almost be symmetric with respect to it. 
On the other hand, as shown in \cite{UCO_aps} and as can also be seen by comparing 
Eqs. (\ref{eq:strong_alpha_1})-(\ref{eq:strong_bbar_1}) with Eqs. (\ref{eq:strong_alpha_2})-(\ref{eq:strong_bbar_2}), 
the relativistic images that form due a photon sphere from $ b>b_{th} $ side and $ b<b_{th} $ side are not symmetric 
with respect to it. We shall see in Sec. \ref{sec:observables} that 
in case of the images formed due to the photon sphere located outside the throat, the images formed inside the 
photon sphere ($b<b_m$) will have much larger angular separation and magnification than those formed outside it ($b>b_m$).


\section{Examples}
\label{sec:examples}
\subsection{Strong lensing by Ellis-Bronnikov wormhole}

The line element of the Ellis-Bronnikov wormhole \cite{ellis1,ellis2}, in Schwarzschild radial coordinate, is given by
\begin{equation}
ds^2=-dt^2 + \frac{dr^2}{1-\frac{r_{th}^2}{r^2}} + r^2 d\Omega^2,
\label{eq:Ellis-Bronikov1}
\end{equation}
or, can be written in the proper radial coordinate as
\begin{equation}
ds^2=-dt^2 + dl^2 + \left(l^2 + r_{th}^2 \right) d\Omega^2.
\label{eq:Ellis-Bronikov2}
\end{equation}
Comparing Eqs. (\ref{eq:Ellis-Bronikov1}) and (\ref{genmetric}), we find
\begin{equation}
A(r)=1, \quad B(r)= \frac{1}{1-\frac{r_{th}^2}{r^2}}, \quad C(r)=r^2.
\end{equation}
The wormhole throat in this case acts as an effective photon sphere, as can been seen from the effective 
potential $V_{eff}=L^2/(l^2+r_{th}^2)$ in the proper radial coordinate. The corresponding critical impact parameter 
is $ b_{th}=\sqrt{\frac{C(r_{th})}{A(r_{th})}}=r_{th}$. The bending angle of light in the strong deflection limit in the 
background of this wormhole spacetime is analyzed below.

\begin{itemize}
\item Approaching strong deflection limit from the $ b>b_{th} $ side:\\
In this case, the observer and the light source are on the same side of the throat, and there will be a 
turning point ($ r_0 $) of light  outside the throat ($ r_{th} $). Therefore, the bending angle of light takes the following form:
\begin{equation}
\alpha(r_0)= 2 \int_{r_0}^{\infty} \frac{b(r_0) ~ dr}{r^2 \sqrt{1-\frac{r_{th}^2}{r^2}} \sqrt{1-\frac{b^2}{r^2}}} - \pi,
\end{equation}
where the impact parameter of light $ b $, in terms of the turning point $ r_0 $, is given as $ b(r_0)=\sqrt{\frac{C(r_0)}{A(r_0)}}=r_0 $. 
Hence, the above integration becomes
\begin{equation}
\alpha(r_0)= 2r_0 \int_{r_0}^{\infty} \frac{dr}{r^2 \sqrt{1-\frac{r_{th}^2}{r^2}} \sqrt{1-\frac{r_0^2}{r^2}}} - \pi.
\end{equation}
This integration can be performed exactly. Putting $ \frac{r_0}{r}=\sin y $, we get
\begin{equation}
\alpha(r_0)= 2 \int_{0}^{\pi/2} \frac{dy}{\sqrt{1-\frac{r_{th}^2}{r_0^2}\sin^2 y}} - \pi = 2 K(m) - \pi,
\end{equation}
where $ K(m) $ is the complete elliptic integral of the first kind and $ 0<m<1 $, with $ m=\frac{r_{th}}{r_0} $. The expansion of $ K(m) $ in the limit $m\to 1$ is given by (see Eq. (10) of Sec. 13.8 in \cite{K_book})
\begin{equation}
\lim_{m\to1} K(m)=-\frac{1}{2}\log\left(1-m^2\right)+2\log2 + \mathcal{O}\left[\left(1-m^2\right)\log \left(1-m^2\right)\right].
\label{eq:K_expansion}
\end{equation}
Therefore, in the strong deflection limit, $ r_0 \to r_{th} $ or $ b \to b_{th} $ ($ b\geq b_{th} $), the bending angle can be written as
\begin{equation}
\alpha(b)=-\log\left(\frac{b^2}{b_{th}^2}-1\right) + 4\log2 - \pi + \mathcal{O}\left[\left(b^2-b_{th}^2\right)\log \left(b^2-b_{th}^2\right)\right].
\label{eq:alpha-ellis-exact}
\end{equation}
Let us find out the same bending angle in the strong deflection limit by the line element in Eq. (\ref{eq:Ellis-Bronikov1}), 
using the expressions we derived in Eqs. (\ref{eq:strong_alpha_3})-(\ref{eq:strong_bbar_3}) directly. Recall once again 
that, in this case, $ A(r)=1 $, $ B(r)= 1/(1-\frac{r_{th}^2}{r^2}) $, or, $ \bar{B}(r)=B(r)^{-1}=1-\frac{r_{th}^2}{r^2} $, and $ C(r)=r^2 $. 
Therefore, $ A_{th}=1 $, $ A'_{th}=0 $, $\bar{B}_{th}=0 $, $ \bar{B}'_{th}=\frac{2}{r_{th}} $, $ C_{th}=r_{th}^2 $, and $ C'_{th}=2r_{th} $. 
Using these, after some simplifications, the regular part $I_R(r_{th})$ becomes (see Sec. \ref{subsec:sameside_b} for the regular part)
\begin{equation}
I_R(r_{th})=\int_0^1 \frac{dz}{2-z}=\log2.
\end{equation}
Therefore, the expressions of $ \bar{a} $, $ \bar{b} $, and the corresponding deflection angle $ \alpha(b) $ become
\begin{equation}
\bar{a}=1, \quad \bar{b}= 4\log2 - \pi,
\end{equation}
\begin{equation}
\alpha(b)=-\log\left(\frac{b^2}{b_{th}^2}-1\right) + 4\log2 - \pi + \mathcal{O}\left[\left(b^2-b_{th}^2\right)\log \left(b^2-b_{th}^2\right)\right]
\label{eq:alpha-ellis-apprx}
\end{equation}
As expected, Eqs. (\ref{eq:alpha-ellis-exact}) and (\ref{eq:alpha-ellis-apprx}) exactly match with each other. Moreover, the above expression of $ \alpha(b) $ can be further approximated in the limit $b\to b_{th}$ as
\begin{equation}
\alpha(b)\simeq -\log\left[\left(\frac{b}{b_{th}}-1\right)\left(\frac{b}{b_{th}}+1\right)\right] + 4\log2 - \pi \simeq  -\log\left(\frac{b}{b_{th}}-1\right) + 3\log2 - \pi
\label{eq:alpha-ellis-tsuka}
\end{equation}
The above expression for the bending angle is obtained when the wormhole metric is written in spherical polar coordinate system. 
However, the same expression can be obtained when the metric is written in the proper radial coordinate as well.
This has been done in Sec. III(C) of \cite{Tsu1}. Therefore, we can arrive at the bending angle formula in the strong deflection
limit when the metric is expressed either in 
spherical polar coordinate ($r$) or in the proper radial coordinate ($l$). 

The advantage of our bending angle formula 
obtained in Sec. \ref{subsec:sameside_b} is that it does not require any coordinate transformation to write the metric in the proper radial coordinate. Therefore, if the metric of a certain wormhole spacetime cannot be expressed explicitly in the proper radial coordinate, then we cannot apply the method described in Sec. III(C) of \cite{Tsu1}. Instead, the methodology described in this paper can be directly used to obtain the required bending angle.
		
\item Approaching strong deflection limit from $ b<b_{th} $ side:\\
In this case, the observer and the light source are on opposite sides of the throat, and there exists no turning point. This is the case discussed in Sec. \ref{subsec:otherside_b}. The formula for bending angle of light is written as,
\begin{equation}
\alpha(b)=2 \int_{r_{th}}^{\infty} \frac{b ~ dr}{r^2 \sqrt{1-\frac{b_{th}^2}{r^2}} \sqrt{1-\frac{b^2}{r^2}}} - \pi,
\end{equation}
where we have replaced $r_{th}=b_{th}$. Similar to the previous case, assuming $ \frac{b_{th}}{r}=\sin y $, we obtain
\begin{equation}
\alpha(b)= \frac{2b}{b_{th}} \int_{0}^{\pi/2} \frac{dr}{\sqrt{1-\frac{b^2}{b_{th}^2}\sin^2 y}} - \pi = \frac{2b}{b_{th}} K(n) - \pi,
\end{equation}
where, $ n= \frac{b}{b_{th}} $ and $ 0<n<1 $. Therefore, using the expansion (\ref{eq:K_expansion}), the bending angle $\alpha(b)$ in the strong deflection limit $ b\to b_{th} $ ($b\leq b_{th}$) can be written as
\begin{equation}
\alpha(b)=-\log\left(\frac{b_{th}^2}{b^2}-1\right) + 4\log2 - \pi + \mathcal{O}\left[\left(b_{th}^2-b^2\right)\log \left(b_{th}^2-b^2\right)\right].
\label{eq:alpha-ellis-otherside-exact}
\end{equation}
We now obtain the strong bending formula using our analytic formula (\ref{eq:strong_alpha_5_app})-(\ref{eq:strong_bbar_5}) and the metric (\ref{eq:Ellis-Bronikov1}). To this end, we first note that the regular part $I_R(b_{th})$ is given by
\begin{equation}
I_R(b_{th})=\int_0^1 \frac{dz}{2-z}=\log2.
\end{equation}
Therefore, we obtain $\bar{a}=1$, $\bar{b}= 4\log2 - \pi$ and
\begin{equation}
\alpha(b)=-\log\left(\frac{b_{th}^2}{b^2}-1\right) + 4\log2 - \pi + \mathcal{O}\left[\left(b_{th}^2-b^2\right)\log \left(b_{th}^2-b^2\right)\right].
\label{eq:alpha-ellis-otherside-approx}
\end{equation}
Note that, similar to the previous case, in this case also, the strong bending angle (\ref{eq:alpha-ellis-otherside-approx}) matches with that in Eq. (\ref{eq:alpha-ellis-otherside-exact}).
		
\end{itemize}

\subsection{Strong lensing by a wormhole with exponential redshift function}
\begin{figure}[ht]
\centering
\subfigure[]{\includegraphics[scale=0.55]{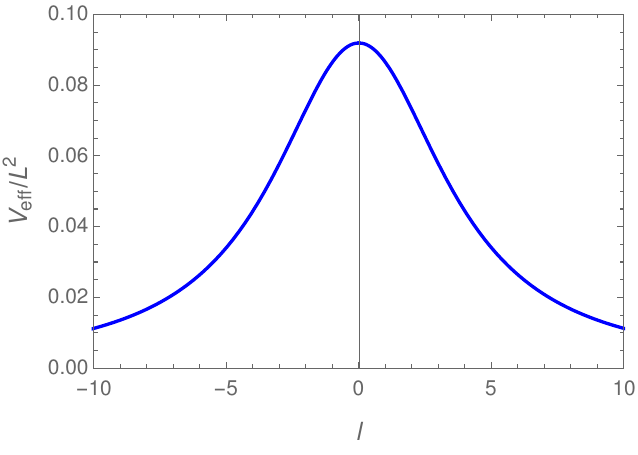}}
\subfigure[]{\includegraphics[scale=0.55]{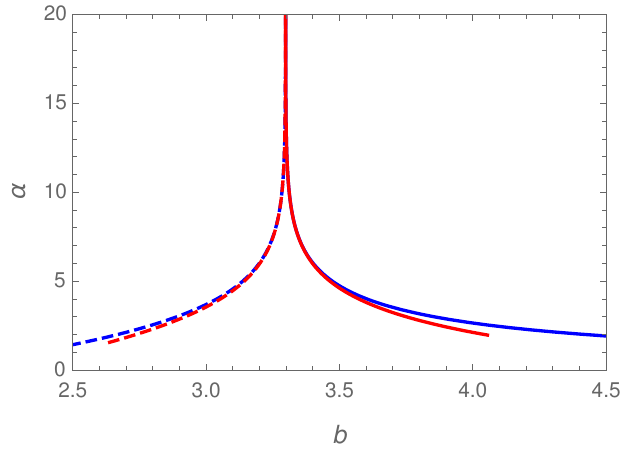}}
\subfigure[]{\includegraphics[scale=0.55]{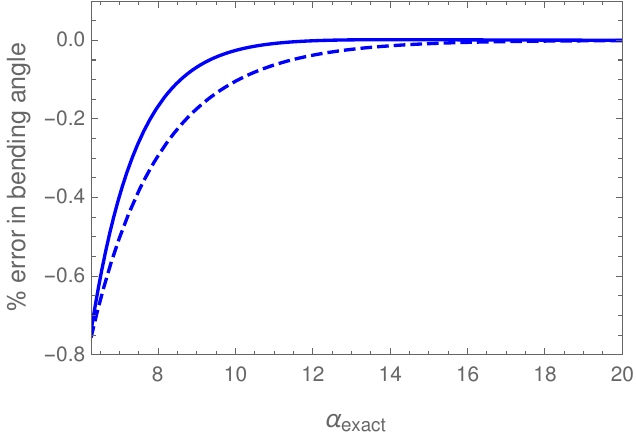}}
\caption{Plots showing (a) the effective potential plotted in proper radial coordinate, (b) numerically integrated (blue curves) and analytic (red curves) deflection angle of light coming from the other side (dashed curves) as well as of that coming from a source located on the observer's side (solid curves) of the throat and (c) $\%$ error $(\frac{\alpha-\alpha_{exact}}{\alpha_{exact}}\times 100)$ in bending angle as a function of $\alpha_{exact}$ in the strong deflection limit $(\alpha\geq 2\pi)$ for light coming from the other side (dashed curve) as well as of that coming from a source located on the observer's side (solid curve) of the throat. Here, $\alpha_{exact}$ is the numerically integrated bending angle and $\alpha$ is the analytic bending angle which we obtained. Here, we have taken $M=1$.}
\label{fig:exp}
\end{figure}

We now consider the following wormhole with a exponential redshift function:
\begin{equation}
ds^2=-e^{-\frac{r_{th}}{r}}dt^2+\frac{dr^2}{1-\frac{r_{th}}{r}}+r^2\left(d\theta^2+\sin^2\theta d\phi^2\right).
\label{eq:exp_metric}
\end{equation}
This type of metric has been frequently used in the literature \cite{WS2,WS3,WS4,WA1}. We can define $r_{th}=2M$, where $M$ is the mass of the wormhole. It can be noted that, unlike the Ellis-Bronnikov wormhole, the above wormhole spacetime can not be explicitly written in the proper radial coordinate, even though the integration in (\ref{eq:proper_radial}) in this case can be performed analytically. Figure \ref{fig:exp} shows the plots for the effective potential plotted in the proper radial coordinate, the deflection angle and the percentage error in the analytic deflection angle. Note that the throat acts as an effective photon sphere in this case. The corresponding critical impact parameter is $b_{th}=r_{th}\sqrt{e}$. For the analytic deflection angle in this case, we have used Eqs. (\ref{eq:strong_alpha_3a})-(\ref{eq:strong_bbar_3a}) when the light source is on the observer's side and Eqs. (\ref{eq:strong_alpha_5_app})-(\ref{eq:strong_bbar_5}) when the light source is on the other side of the throat.

\subsection{Strong lensing by a wormhole with vanishing curvature}

\begin{figure}[ht]
\centering
\subfigure[]{\includegraphics[scale=0.55]{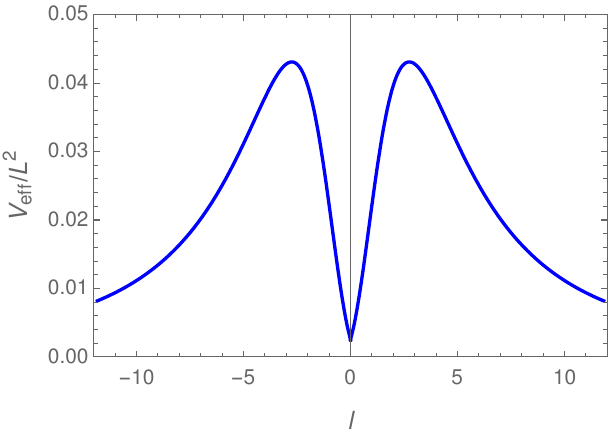}}
\subfigure[]{\includegraphics[scale=0.55]{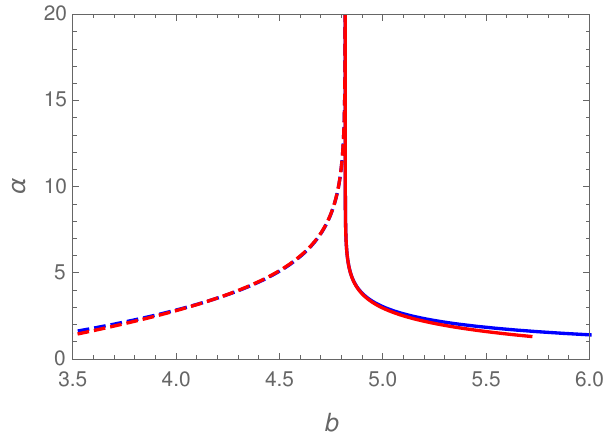}}
\subfigure[]{\includegraphics[scale=0.55]{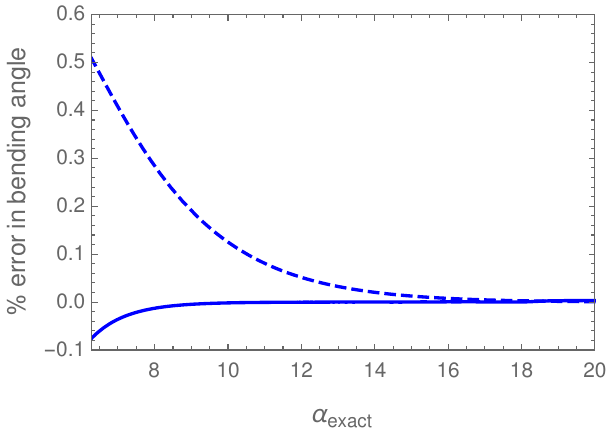}}
\caption{Plots showing (a) the effective potential plotted in proper radial coordinate, (b) numerically integrated (blue curves) and analytic (red curves) deflection angle of light coming from the other side (dashed curves) as well as of that coming from a source located on the observer's side (solid curves) of the throat and (c) $\%$ error $(\frac{\alpha-\alpha_{exact}}{\alpha_{exact}}\times 100)$ in bending angle as a function of $\alpha_{exact}$ in the strong deflection limit $(\alpha\geq 2\pi)$ for light coming from the other side (dashed curve) as well as of that coming from a source located on the observer's side (solid curve) of the throat. Here, $\alpha_{exact}$ is the numerically integrated bending angle and $\alpha$ is the analytic bending angle which we obtained. Here, we have taken $M=1$.}
\label{fig:ST}
\end{figure}

Next, we consider strong lensing by the wormhole \cite{kappa1,kappa2}
\begin{equation}
ds^2=-\left(\kappa+\lambda \sqrt{1-\frac{2M}{r}} \right)^2 dt^2+\frac{dr^2}{1-\frac{2M}{r}}+r^2\left(d\theta^2+\sin^2\theta d\phi^2\right)
\label{eq:ST_metric}
\end{equation}
which has vanishing Ricci scalar. Here, we choose the parameter $\kappa$ and $\lambda$ to be positive in such a way that $(\kappa+\lambda)=1$. This is to ensure that the metric function $|g_{tt}|\to 1$ as $r\to \infty$. When $\kappa=0$, we obtain the Schwarzschild black hole with mass $M$. The throat is at $r=r_{th}=2M$. The above wormhole has a photon sphere at $r=r_m$ on each side of the throat, where
\begin{equation}
r_m=\frac{2M}{1-\left(\frac{\sqrt{\kappa^2+3\lambda^2}-\kappa}{3\lambda}\right)^2}\geq r_{th}.
\end{equation}
Note that, when $\kappa=0$, we recover the photon sphere radius $r_m=3M$ of the Schwarzschild black hole. Figure \ref{fig:ST} shows the plots for the effective potential plotted in the proper radial coordinate, the deflection angle and the percentage error in the analytic deflection angle. For the analytic deflection angle in this case, we have used Eqs. (\ref{eq:strong_alpha_1})-(\ref{eq:strong_bbar_1}) when the light source is on the observer's side and Eqs. (\ref{eq:strong_alpha_4})-(\ref{eq:strong_bbar_4}) when the light source is on the other side of the throat.

\section{Observables in gravitational lensing}
\label{sec:observables}

We now discuss various observables of the relativistic images formed due to the strong gravitational lensing. We assume that the observer and the light source are faraway from the wormhole throat. It is to be noted that, when the observer and the light source are on the same side of the throat, the relativistic images due to strong lensing by a photon sphere discussed in Sec. \ref{subsec:sameside_a} and that by a throat discussed in Sec. \ref{subsec:sameside_b} are formed at impact parameters greater than the corresponding critical value, i.e., they are formed outside the photon sphere and the throat respectively. The dependence of the strong deflection angle on $b$ and $b_{m}$ or $b_{th}$ in these two cases are similar to each other [see Eqs. (\ref{eq:strong_alpha_1})-(\ref{eq:strong_bbar_1}) and Eqs. (\ref{eq:strong_alpha_3a})-(\ref{eq:strong_bbar_3a})], except that the expressions for $\bar{a}$ and $\bar{b}$ are different in the two cases. 

Also note that, as we discussed in \ref{subsec:sameside_a}, the expression for the strong bending angle due to the photon sphere of a wormhole is the same as that due to the photon sphere of a black hole. Therefore, the expressions for the angular position and magnification of the relativistic images formed in these two wormhole cases will be the same as those due to a black hole and for the $n$th relativistic image, are, respectively, given by \cite{SL3}
\begin{equation}
\theta_n=\frac{b_{m,th}}{D_{OL}}(1+e_n)=\theta_{\infty}(1+e_n), \quad e_n=e^{\frac{\bar{b}-2n\pi}{\bar{a}}},
\end{equation}
\begin{equation}
\mu_n = \frac{b_{m,th}^2 D_{OS}e_n(1+e_n)}{\bar{a}\beta D_{OL}^2 D_{LS}},
\end{equation}
where $b_{m,th}$ means either $b_m$ or $b_{th}$, and $\theta_{\infty}=b_{m,th}/D_{OL}$ is the angular position of the relativistic image formed either at the photon sphere or at the throat. Depending on the cases, we use the corresponding expressions for $\bar{a}$ and $\bar{b}$ in the above equations. Here, $D_{LS}$ is the distance between the lens and the source, $D_{OS}$ is the distance between the observer and the source, $D_{OS}=D_{OL}+D_{LS}$, $D_{OL}$ is the distance between the observer and the lens, $\beta$ is the angular 
separation between the source and the lens. Note that the angular position of the images decreases with $n$, implying that, in the images formed outside the photon sphere or the throat, the first relativistic image is the outermost one and the image with the angular position $\theta_{\infty}$ is the innermost one. Moreover, we can define another observable, namely the angular separation $s_n$ between the $n$th and $(n+1)$th images as
\begin{equation}
s_n=\theta_n-\theta_{n+1}.
\end{equation}

However, when the observer and the light source are on the same side of the throat and the strong lensing takes place due to the presence of an antiphoton sphere as discussed in Sec. \ref{subsec:sameside_c}, the images are formed at impact parameter less than the critical value $b_m$, i.e., they are formed inside the photon sphere. Note that, as discussed in Sec. \ref{subsec:sameside_c}, the analytic strong deflection formula in this case is the same as that due to the presence of an antiphoton sphere around an ultracompact object \cite{UCO_aps}. Therefore, the expressions for the angular position and magnification of the $n$th relativistic image in this case are given by \cite{UCO_aps}
\begin{equation}
\theta_{-n}=\frac{b_m}{D_{OL}}\frac{1}{\sqrt{1+e_{-n}}}=\frac{\theta_{-\infty}}{\sqrt{1+e_{-n}}}, \quad e_{-n}=e^{\frac{\bar{b}-2n\pi}{\bar{a}}},
\label{eq:ap_UCO}
\end{equation}
\begin{equation}
\mu_{-n} = -\frac{b_m^2 D_{OS}}{2\bar{a}\beta D_{OL}^2 D_{LS}}\frac{e_{-n}}{(1+e_{-n})^2},
\label{eq:mag_UCO}
\end{equation}
where $\theta_{-\infty}=b_m/D_{OL}$ is the angular position of the relativistic image formed at the photon sphere. Here, the minus sign before $n$ implies that the images are formed inside the photon sphere. Note that, in contrast to that for the images formed outside the photon sphere or the throat (previous two cases), the angular position of the images formed inside the photon sphere increases with $n$, implying that, in the these inner images, the first relativistic image is the innermost one and the image with the angular position $\theta_{-\infty}$ is the outermost one.

In the scenario discussed in Sec. \ref{sec:otherside}, i.e., when the observer and the light source are on the opposite sides of the throat, the images are formed at the impact parameters smaller than the critical value, i.e., they are formed either inside the photon sphere (Sec. \ref{subsec:otherside_a}) or inside the throat (Sec. \ref{subsec:otherside_b}). Note that the dependence of the strong bending angle on $b$ and $b_m$ or $b_{th}$ in these cases [see Eqs. (\ref{eq:strong_alpha_4})-(\ref{eq:strong_bbar_4}) and Eqs. (\ref{eq:strong_alpha_5_app})-(\ref{eq:strong_bbar_5})] is similar to the one obtained in the strong lensing due to the antiphoton sphere [see Eqs. (\ref{eq:strong_alpha_2})-(\ref{eq:strong_bbar_2})], except that the expression for $\bar{a}$ and $\bar{b}$ are different. Therefore, 
when the observer and the light source are on the opposite sides of the throat also, the angular positions and magnifications of the relativistic images are given by Eqs. (\ref{eq:ap_UCO}) and (\ref{eq:mag_UCO}) with $b_m$ replaced by $b_{m,th}$ and $\bar{a}$ and $\bar{b}$ given by the corresponding expressions. Note that $\theta_{\infty}=\theta_{-\infty}$. Beside the angular positions and magnifications of the relativistic images formed in these cases, we define one more observable, namely the angular separation $s_{-n}$ between the $n$th and $(n+1)$th images as
\begin{equation}
s_{-n}=\vert\theta_{-n}-\theta_{-(n+1)}\vert.
\end{equation}

The angular positions, angular separations and the magnifications of the relativistic images of the examples discussed in the previous section along with the Schwarzschild black hole are presented in Table \ref{Table1}. Here we have restored $G$ and $c$ by replacing $M$ by $(GM)/c^2$. Here, the parameter $M$ and the distance $D_{OL}$ are, respectively, taken to be equal to the mass and distance of the supermassive black hole Sgr A$^*$ at center of our Galaxy. Although the Ellis-Bronnikov wormhole is massless, we have taken its throat size $r_{th}$ to be equal to $2M$ for simplicity.
\begin{table}[h!]
\centering
\caption{The angles are in microarc sec. Here, we have taken $M=4.31\times 10^6 M_{\odot}$, $D_{OL}=7.86$ Kpc, which are the parameters for the supermassive black hole Sgr A$^*$ at center of our Galaxy, $D_{LS}=D_{OL}$ and $\beta=5^\circ$. For the Ellis-Bronnikov wormhole, the throat size $r_{th}$ is taken to be equal to $2M$.}
\begin{tabular}{| c | c | c | c | c |} 
\hline\hline
& Schwarzschild & Ellis- & Wormhole & Wormhole \\
  & black hole & Bronnikov & in & in Eq. (\ref{eq:ST_metric}) \\
  & & wormhole & Eq. (\ref{eq:exp_metric}) & $\kappa=0.1$, $\lambda=0.9$ \\
 \hline
  $\theta_1$ & 28.2802 & 10.9476 & 18.4010 & 26.2277 \\
  $\theta_2$ & 28.2449 & 10.8723 & 17.9446 & 26.1884 \\
  $\theta_{\infty}$ & 28.2449 & 10.8715 & 17.9240 & 26.1883 \\
  $\theta_{-2}$ & --- & 10.8706 & 17.9034 & 26.1363 \\
  $\theta_{-1}$ & --- & 10.7961 & 17.4652 & 25.1555 \\
  $\mu_1\times 10^{22}$ & 5.3850 & 3.1750 & 23.6420 & 5.3934 \\
  $\mu_{2}\times 10^{22}$ & 0.0100 & 0.0371 & 0.9963 & 0.0122 \\
  $\mu_{-2}\times 10^{22}$ & --- & $-0.0371$ & $-0.9906$ & $-3.5407$ \\
  $\mu_{-1}\times 10^{22}$ & --- & $-3.0664$ & $-20.7603$ & $-63.8772$ \\
  $s_1$ & 0.0353 & 0.0752 & 0.4564 & 0.0393 \\
  $s_{2}$ & 0.0001 & 0.0009 & 0.0197 & 0.0381 \\
  $s_{-2}$ & --- & 0.0009 & 0.0197 & 0.4208 \\
  $s_{-1}$ & --- & 0.0745 & 0.4382 & 0.9807 \\
 \hline\hline
 \end{tabular}
\label{Table1}
\end{table}

Note that the angular separations and magnifications of the relativistic images formed due to the strong lensing by the throat are almost symmetric with respect to the throat (see third and fourth columns of Table \ref{Table1}). The reason for this has been discussed already at the end of Sec. \ref{subsec:otherside_b}. On the other hand, when the images are formed due to the strong lensing by a photon sphere located outside the throat, their angular separations and magnifications are asymmetric with respect to the photon sphere. The images formed inside the photon sphere have much larger angular separations and magnification than those formed outside it (see fifth column of Table \ref{Table1}). Therefore, although both the throat and the photon sphere act as the maxima of the effective potential for image formation, the unique characteristic feature of symmetry of images in case of a wormhole throat distinguishes it from usual photon sphere. This property is very interesting and may serve as a significant tool in futuristic experiments regarding wormholes.

We recall that, for black holes, relativistic images are always formed outside the photon spheres. 
Whereas, for wormholes, when the strong lensing takes place due to the presence of an antiphoton sphere (case {\bf a3}), 
the images are formed both inside and outside the photon sphere simultaneously, even if the observer and the source are 
on the same side of the throat. Except this case, in all other cases, the images are formed both inside and outside a photon 
sphere or a throat simultaneously only when there are light sources present on both sides, i.e., on the observer's side as 
well as on the opposite side of the throat. If a light source is present only on one side, then, except in the case {\bf a3}, 
the images are formed either the inside or the outside of the throat or the photon sphere, depending on which side 
the source is present. 

Therefore, when the observer and the source are on opposite sides of the throat, the relativistic images are formed 
only inside the photon sphere (case {\bf b1}) or the throat (case {\bf b2}). Note that the angular separation and 
magnification of these images formed only inside the photon sphere or the throat decrease as we move from the 
innermost (first) image to the outermost one (see Table \ref{Table1}). Whereas, for black holes, it's the opposite. 
The images for black holes are always formed outside the photon sphere and their angular separation and 
magnification decrease as we move from the outermost (first) image to the innermost one. Therefore, this unique 
lensing feature of the images formed only inside the photon sphere or the throat due to the strong lensing of light coming from the other side of the 
throat (cases {\bf b1} and {\bf b2}) can help us detecting wormholes in futuristic experiments.


\section{Summary and conclusions}
\label{sec:conclusions}

In this paper, we have carried out an exhaustive analysis of gravitational lensing in the strong field limit from wormholes. 
We have classified and studied five different possibilities that can arise, and exemplified our computations (carried out for 
generic static spherically symmetric wormhole spacetimes) with three distinct examples. We have pointed out several 
distinctive features of strong lensing from wormholes as compared to those from black holes. For black holes, relativistic
 images are always formed outside the photon spheres, and their angular separation and magnification decrease as we 
 move from the outermost (first) image to the innermost one formed at the photon sphere. 
 
In contrast, for wormholes,
images can be formed both inside and outside the photon spheres or the throat, and the angular separation and 
magnification of the former (i.e. images formed inside the photon sphere or the throat) are opposite in nature as 
compared to the latter (i.e. images formed outside the photon sphere or the throat) or as compared to those by 
black holes, i.e., they decrease as we move from the innermost (first) image to the outermost one formed at the 
photon sphere. Depending on the situations, the relativistic images formed due to strong lensing by wormholes
can have the following patterns:

\begin{itemize}
\item The images are formed only outside the photon sphere (case {\bf a1}) or the throat (case {\bf a2}) when the light source is present only on the observer's side of the throat. Qualitatively, this is very similar to that by a black hole. However, the strong bending formula which we obtained in case {\bf a2} is different from that in case of a black hole.

\item The images are formed both outside and inside the photon sphere when the light source is present only on the observer's side of the throat and the strong lensing takes place in the presence of an antiphoton sphere (case {\bf a3}). Qualitatively, this is very similar to that by other horizonless ultracompact objects discussed in \cite{UCO_aps}.

\item The images are formed both outside and inside the photon sphere (a combination of the cases {\bf a1} and {\bf b1}) or the throat (a combination of the cases {\bf a2} and {\bf b2}) when there are light sources present on both sides of the throat. As we have discussed in the previous section, the images formed due to a photon sphere in this case are asymmetric with respect to it, whereas those formed due to a throat are almost symmetric with respect to it.

\item The images are formed only inside the photon sphere (case {\bf b1}) or the throat (case {\bf b2}) when the observer and the light source are on opposite sides of the throat. This case is qualitatively very different from that by a black hole.

\end{itemize}

Note once again that as we have discussed above, the angular separation and magnification of the images formed inside the photon sphere or the throat in different strong lensing cases by the wormhole are opposite in nature as compared to the ones by the black holes. This fact together with the above patterns of the images formed provides several distinctive features of strong lensing from wormholes as compared to those from black holes. These distinctive features may be useful to detect wormholes in futuristic experiments. 

An immediate interesting extension of the analysis presented here would be to compute strong lensing from rotating wormholes. We leave this 
for a future publication.


\appendix
\renewcommand{\theequation}{A.\arabic{equation}}
\setcounter{equation}{0}
\section{Details of the calculation for case {\bf b2}}
\label{sec:Appendix}

To obtain the strong deflection angle in this case, we define
\begin{equation}
	z= 1-\frac{r_{th}}{r},
\end{equation}
Putting this in Eq. (\ref{eq:alpha_b}), we obtain
\begin{equation}
	I(b)=\int^{1}_{0}f(z,b,r_{th})dz,
\end{equation}
where
\begin{equation}
	f(z,b,r_{th})= \frac{2r_{th}}{\sqrt{G(z,b,r_{th})}}~, \quad
	G(z,b,r_{th})= R\frac{C}{B}(1-z)^{4}
\end{equation}
Therefore, $R(r)$ can be expanded in the power of $z$ as
\begin{align}
    R(r) =&\left(\frac{C_{th}}{b^2 A_{th}}-1\right) +\frac{C_{th}}{b^2 A_{th}} \left[ r_{th}\left(\frac{C_{th}^{'}}{C_{th}}-\frac{A_{th}^{'}}{A_{th}}\right)z \right. \nonumber \\
    & \left. 
    + \Bigg\{\frac{r^2_{th}}{2}\left(\frac{C_{th}^{''}}{C_{th}}-\frac{A_{th}^{''}}{A_{th}}\right) + r_{th}\left(1-\frac{A^{'}_{th} r_{th}}{A_{th}}\right)\left(\frac{C_{th}^{'}}{C_{th}}-\frac{A_{th}^{'}}{A_{th}}\right)\Bigg\}z^2\right] +\mathcal{O}(z^3).
\end{align}
 Once again, we define $\bar{B}(r)=B(r)^{-1}$ so that, by definition of the throat, $ \bar{B}(r_{th})=\bar{B}_{th}=0 $. Therefore, the expansion of $\bar{B}(r)$ in powers of $ z $ will take the form,
\begin{equation}
	\bar{B}(r)=\bar{B}^{'}_{th}r_{th}z +\left(\frac{1}{2}\bar{B}^{''}_{th}r^2_{th}+\bar{B}^{'}_{th}r_{th}\right)z^2 + \mathcal{O}(z^3)
\end{equation}
Using similar expansion of $ C $ in addition to $ \bar{B} $ and $ R $, we obtain the expansion of $G(z,b,r_{th})$ in powers of $z$ as 
\begin{equation}
	G(z,b,r_{th})=\delta z+\eta z^2+\mathcal{O}(z^3),
\end{equation}
where
\begin{equation}
	\delta=C_{th}\bar{B}^{'}_{th}r_{th}\left(\frac{C_{th}}{b^2 A_{th}}-1\right),
\end{equation}

\begin{align}
	\eta = r_{th}\left(\frac{C_{th}}{b^2 A_{th}}-1\right)\left(\frac{1}{2}C_{th}\bar{B}^{''}_{th}r_{th}-3C_{th}\bar{B}^{'}_{th}+C^{'}_{th}\bar{B}^{'}_{th}r_{th}\right)+r^2_{th}\frac{C^2_{th}\bar{B}^{'}_{th}}{b^2 A_{th}}\left(\frac{C^{'}_{th}}{C_{th}}-\frac{A^{'}_{th}}{A_{th}}\right)
\end{align}
Note that $\frac{C(r_{th})}{A(r_{th})}=b_{th}^2$. Therefore, in the limit $b \to b_{th}$, $\left(\frac{C_{th}}{b^2 A_{th}}-1\right)\to 0$. As a result, in this limit, we obtain
\begin{equation}
\delta_{th}=\delta\vert_{b=b_{th}}=0, \quad \eta_{th}=\eta\vert_{b=b_{th}}=r^2_{th}C_{th}\bar{B}^{'}_{th}\left(\frac{C^{'}_{th}}{C_{th}}-\frac{A^{'}_{th}}{A_{th}}\right).
\end{equation}
Hence, we obtain
\begin{equation}
	G_{th}(z)=\eta_{th} z^{2}+\mathcal{O}(z^{3}).
\end{equation}
This again shows that the leading order of the divergence of $f(z, b,r_{th})$ is $z^{-1}$ and that the integral $I(b)$ diverges logarithmically in the strong deflection limit $b\to b_{th}$.

As usual, we again split up the integral $I(b)$ into a divergent part $I_D(b)$ and a regular part $I_R(b)$ giving, $I(b)=I_{D}(b)+I_{R}(b)$. The divergent part $I_D(b)$ is given by
\begin{equation}
	I_D(b)= \int^{1}_{0}f_D(z,b,r_{th})dz~, \quad
	f_D(z,b,r_{th})=\frac{2r_{th}}{\sqrt{\delta z+\eta z^2}}
\end{equation}
And the regular part $I_R(b)$ is defined as
\begin{equation}
	I_{R}(b)=\int^{1}_{0} f_R(z,b,r_{th})dz~, \quad
	f_R(z,b,r_{th})=f(z,b,r_{th})-f_D(z,b,r_{th})
\end{equation}
After performing the integration, the divergent part $I_D(b)$ becomes
\begin{equation}
	I_D(b)=\frac{4r_{th}}{\sqrt{\eta}}\log \frac{\sqrt{\eta}+\sqrt{\delta+\eta}}{\sqrt{\delta}}
\end{equation}
Therefore, in the limit $b\to b_{th}$, we get
\begin{equation}
	I_D(b)=-\frac{2r_{th}}{\sqrt{\eta_{th}}}\log \left(\frac{b^2_{th}}{b^2}-1\right) + \frac{2r_{th}}{\sqrt{\eta_{th}}}\log \left[4r_{th}\left(\frac{C^{'}_{th}}{C_{th}}-\frac{A^{'}_{th}}{A_{th}}\right)\right] + \mathcal{O}\left[\left(b^2_{th}-b^2\right)\log\left(b^2_{th}-b^2\right)\right]
\end{equation}
The corresponding expansion of the regular part $I_R(b)$ in powers of ($b_{th}^2-b^2$) in the strong deflection limit $b\to b_{th}$ takes the form,
\begin{equation}
	I_R(b)= \int^{1}_{0} f_R(z,b_{th},r_{th})dz+\mathcal{O}\left[\left(b_{th}^2-b^2\right)\log \left(b_{th}^2-b^2\right)\right]
\end{equation}
Finally, we obtain the expression of deflection angle in the strong deflection limit $b\to b_{th}$ ($b<b_{th}$) as
\begin{equation}
\alpha(b)=-\bar{a}\log \left( \frac{b_{th}^2}{b^2}-1 \right) +\bar{b} + \mathcal{O}\left[\left(b_{th}^2-b^2\right)\log \left(b_{th}^2-b^2\right)\right],
\label{eq:strong_alpha_5}
\end{equation}
where
\begin{equation}
\bar{a}=2\sqrt{\frac{A_{th}}{\bar{B}^{'}_{th}\left(C^{'}_{th}A_{th}-A^{'}_{th}C_{th}\right)}}, \quad \bar{b}=\bar{a}\log \left[4r_{th}\left(\frac{C^{'}_{th}}{C_{th}}-\frac{A^{'}_{th}}{A_{th}}\right)\right] +I_R(b_{th})-\pi.
\end{equation}

\end{document}